\documentstyle[psfig]{mn} 
\newcommand{\HI}{H\,{\sc i}}
\newcommand{\HII}{H\,{\sc ii}}
\newcommand{\kms}{km~s$^{-1}$}
\def\gapp{\ifmmode\stackrel{>}{_{\sim}}\else$\stackrel{>}{_{\sim}}$\fi}
\begin{document}

\title[Multi-epoch \HI\ line measurements of pulsars]
{Multi-epoch \HI\ line measurements of 4 southern pulsars}

\author[Johnston et al.]
{Simon Johnston$^1$, B\"arbel Koribalski$^2$, Warwick Wilson$^2$ \&
Mark Walker$^{1,2}$\\
$^1$School of Physics, University of Sydney, NSW 2006, Australia\\
$^2$Australia Telescope National Facility, CSIRO, Epping, NSW 1710,
Australia}

\maketitle

\begin{abstract}
We have measured 21-cm absorption spectra in the direction of three
southern pulsars, PSRs B0736--40, B1451--68 and B1641--45, in three 
separate epochs spread over 2.5 yr. We see no evidence for any
changes in the absorption spectra over this time span in spite
of good velocity resolution and sensitivity. Towards PSR B1641--45
we place an upper limit of $10^{19}$ cm$^{-2}$ on the change
in the column density of the cold, neutral gas.
In addition, we observed PSR B1557--50 and compared its \HI\ absorption
spectrum with spectra taken in 1980 and 1994. A prominent deep 
absorption feature seen at --110 \kms\ in 1994 is weaker in 2000
and likely was not present at all in 1980. Using the standard
interpretation which links the distance traversed 
by the pulsar to a physical cloud size, this results
in a cloud size of $\sim$1000~AU and density of $2\times 10^4$~cm$^{-3}$,
parameters typical of those seen in other observations.
These results are also consistent with an alternative model by
Deshpande (2000) who expects the largest variations in optical 
depth to be seen against the longest time intervals.
\end{abstract}

\begin{keywords}
 pulsars: general --- radio lines: ISM --- stars: distances
\end{keywords}

\section{Introduction}
In 1990, a major review of \HI\ in the Galaxy found little evidence for 
structure in the interstellar medium (ISM)
on scales less than about $10^{16}$~m \cite{dl90}.
However, subsequent studies have all reached the conclusion
that virtually every line of sight exhibits structure on
scales at least
3 orders of magnitude smaller than this.
Recent interest has focussed on determining structure at the very
small scales (hundreds of AU and below). Different techniques have been used,
(a) looking at the changes in \HI\ and molecular profiles in high-resolution 
images of extragalactic objects across the source and/or over 
a period of time \cite{dgr+89,mm95,ddg96,fgdt98,fg01},
(b) observing {Na\,{\sc i}} absorption lines along
lines of sight to binary stars \cite{wm96,lmb00}, and globular
clusters \cite{ml99} and
(c) observing changes in the \HI\ absorption profile of 
strong pulsars over a period of years \cite{fchw91,dmra92,fwcm94}.
\begin{table*} 
\begin{tabular}{crrccccrrrr}
\hline
 PSR   & \multicolumn{1}{c}{$l$} & \multicolumn{1}{c}{$b$} & DM  &
Period & S$_{\rm 20}$ & resolution & 
T (min) & T (min) & T (min) & T (min) \\
 (B1950) &  &    &(pc\,cm$^{-3}$)& (ms)   & (mJy)
& (km~s$^{-1}$) & 94 Oct & 98 Mar & 98 Sep & 00 Aug \\
\hline
0736--40 & 254\fdg2 &--9\fdg2 & 160.8 & 374.9 & 80
         & 0.12 &   90.0 & 1137.4 & 709.9 & 1354.4 \\
1451--68 & 313\fdg9 &--8\fdg5 &   8.7 & 263.4 & 80
         & 0.25 &  270.0 &  601.8 & 512.4 &  462.7 \\
1557--50 & 330\fdg7 &  1\fdg6 & 262.8 & 192.6 & 20
         & 2.00 &  231.7 &        &       &  367.1 \\
1641--45 & 339\fdg2 &--0\fdg2 & 485.3 & 455.1 & 430
         & 0.25 &        &  812.9 & 865.9 & 1192.4 \\
\hline
\end{tabular}
\caption[]{Pulsar parameters, correlator setup and integration times} 
\end{table*}

For the most part, these observations have been interpreted in
terms of individual gas clouds moving in/out of the line-of-sight;
the small (transverse) length-scales, coupled with the substantial
changes in column-density then imply small, dense clouds whose
properties challenge our understanding of the ISM.
The densities inferred for these small clouds are
typically $\sim 2\times 10^{4}$ cm$^{-3}$, and even at temperatures
as low as 10~K, such clouds are over-pressured with respect to
other phases of the ISM, and are thus
expected to dissipate on a time-scale of $\sim100$ yr,
unless some containment mechanism can be found. Possible mechanisms
include magnetic confinement of the clouds, or that clumpy
molecular clouds may have structure with a high fractal dimension.
Heiles (1997)\nocite{hei97} has proposed a geometrical solution 
in terms of sheets and filaments rather than spherical clouds.
Gwinn (2001)\nocite{gwi01} has a model in which interstellar scintillation
differences on-line compared to off-line can induce the variations
seen in the difference spectra of Frail et al. (1994).
Deshpande (2000)\nocite{desh00} recently provided an alternative 
solution to the problem. Based on observations made towards Cas~A and Cyg~A
\cite{ddg00} he suggests that observers are detecting (small) optical 
depth changes along different lines of sight in large clouds, with
contributions to the difference in column-densities arising
{\em from the whole line-of-sight.} In other words, as Deshpande (2000)
points out, a given transverse length scale does not correspond
to a unique three-dimensional length scale, so that physically
small clouds need not exist at all, thus avoiding any difficulty
with over-pressured clouds. Whether or not the fluctuations seen against
extragalactic sources can be extrapolated down to the few AU level
remains to be settled.
Faison et al. (1998)\nocite{fgdt98}, for example, showed that not all optical
depth variations observed towards extragalactic sources fit this model.
Their \HI\ spectra towards 3C138 show variations
far larger than expected.

Frail et al. (1994)\nocite{fwcm94} observed six pulsars with the
Arecibo telescope over three epochs spanning 1.7 yr with a
velocity resolution of 0.5 \kms. They saw
opacity variations towards many individual
features in all six objects. As the origin of the small-scale
fluctuations is far from
clear cut, and as other authors have urged \cite{hei97,fg01},
we therefore decided to follow the 
experiment of Frail et al. (1994)\nocite{fwcm94} with a similar 
one observing southern pulsars. A new
correlator at the Parkes telescope has enabled us to observe both
with 0.1 \kms\ velocity resolution and sufficient
time resolution to allow pulse gating.

\section{Source Selection}
Our choice of pulsars was essentially dictated by the need to observe
the brightest sources at 1420 MHz, the frequency of the
\HI\ line. Although the nearby Vela pulsar
heads this list, preliminary observations showed that 
there is no \HI\ absorption in its direction and it was excluded. (We
note, also, that the Vela pulsar exhibits deep and variable
spectral modulation on the frequency scales relevant here.
This modulation, which is due to diffractive
interstellar scintillation, increases the variance on the
average spectrum at any epoch to a value much greater than
the thermal noise level.)  This left
PSRs B0736--40, B1451--68 and B1641--45 as the candidate stars.
It so happens that their properties are very different.

PSR B0736--40 was most recently observed in \HI\ by
Johnston et al. (1996)\nocite{jkww96} who obtained a lower distance
limit of 2.1 kpc. The pulsar lies behind the Gum Nebula which
contributes to its dispersion measure but even then the mean electron density
along the line of sight is
large for its distance. Proper motion measurements
\cite{dr83,fgl+92} imply a velocity of 800$\pm$130 \kms\ at the lower
distance limit which indicates that the pulsar is unlikely to be 
much further than this. Mitra \& Ramachandran (2001)\nocite{mr01}
assign a distance of 4.5 kpc to this pulsar based on scattering measurements
but we note that they have not taken into account the \HII\ region
along the line of sight discussed in Johnston et al. (1996)\nocite{jkww96}.
A velocity of 800 \kms\ implies the pulsar moves a distance
of 170~AU~yr$^{-1}$.

PSR B1451--68 has a parallax distance of 450$\pm$60 pc \cite{bmk+90a} and 
was observed in \HI\ by Koribalski et al. (1995)\nocite{kjww95}.
Only local absorption was detected. Its proper motion of 41~mas~yr$^{-1}$
\cite{bmk+90a} translates to a distance across 
our line of sight of 18~AU~yr$^{-1}$.

PSR B1641--45 lies in the Galactic plane at a latitude of
0\fdg2. The most recent \HI\ observations
by Frail et al. (1991)\nocite{fchw91} show deep absorption from
+10 to --59 \kms\ and a lack of absorption at an emission feature at
--77 \kms. The pulsar therefore lies in the distance range 4.2 to 5.0~kpc.

Finally, we added PSR B1557--50 to the list.
Deshpande et al. (1992)\nocite{dmra92} compared \HI\ spectra
from this pulsar taken in
the 1970s and saw differences between them. We had observed this
pulsar previously, in 1994 \cite{jkww01}, and we therefore repeated
the observation again in 2000 August. The lower limit to the distance
from the \HI\ measurements is 6.4~kpc.
Neither PSR B1641--45 nor B1557--50 have measured proper motions.

\section{Observations}
New \HI\ observations towards the southern pulsars PSRs B0736--40,
B1451--68 and B1641--45 were carried out 
on 1998 March 6--9, 1998 September 27--29 and
2000 August 3-9 using the 64-m Parkes radio telescope.
An identical receiver system, correlator backend and
observing procedure were used for all 3 epochs.
However, in 1998, analogue filters were used to achieve a 4 MHz
passband and these were replaced with digital filters prior to 
the 2000 observations.

We used the centre beam of the 13-beam `multibeam' receiver \cite{swb+96}.
The digital correlator was configured to record two 
polarizations each with either 4096 or 8192 channels across a 
total bandwidth of 4 MHz giving a velocity range of $\pm$400 \kms\
and a resolution of 0.25 or 0.12 \kms.
The pulsar period was subdivided into 8 phase bins for the
4096 channel mode or 4 phase bins for the 8192 channel mode.
Data were folded synchronously with the pulsar period for 
an integration time of 10~s. The relative phase of the pulse peak
could be shifted to ensure that the pulse fell entirely within
one phase bin. Prior to the observation, the
local oscillator would be set with the appropriate Doppler delay 
necessary to maintain the \HI\ rest frequency in the same correlator channel.
The total time for any single observations
did not exceed 45~min on any one source in
order to reduce the Doppler smearing to less than one channel during
the integration.

In addition, we made observations of PSR B1557--50 in 2000 August.
Previous observations of this pulsar had been made in October 1994
and an interesting, narrow, deep absorption feature was detected
against gas with a temperature of only 20~K \cite{jkww01}.
The 1994 observations were made with a 512 channel correlator configuration
(velocity resolution 2.0 \kms)
and we repeated this configuration again in 2000.

Table~1 lists the pulsar parameters including the
Galactic coordinates, the dispersion measure,
period and flux density at 1.4 GHz. This information was obtained from 
the catalogue of Taylor, Manchester \& Lyne (1993)\nocite{tml93}.
Table~1 further lists the velocity resolution followed
by the observing time in minutes during the 1994 October, 1998 March, 
1998 September and 2000 August observing sessions.
The 1994 October observations of PSRs B0736--40 and B1451--68
are described in Johnston et al. (1996)\nocite{jkww96} and 
Koribalski et al. (1995)\nocite{kjww95}. The velocity resolution of these
observations was 2.0 \kms.

\section{Data Analysis}
\begin{figure}
  \centerline{\psfig{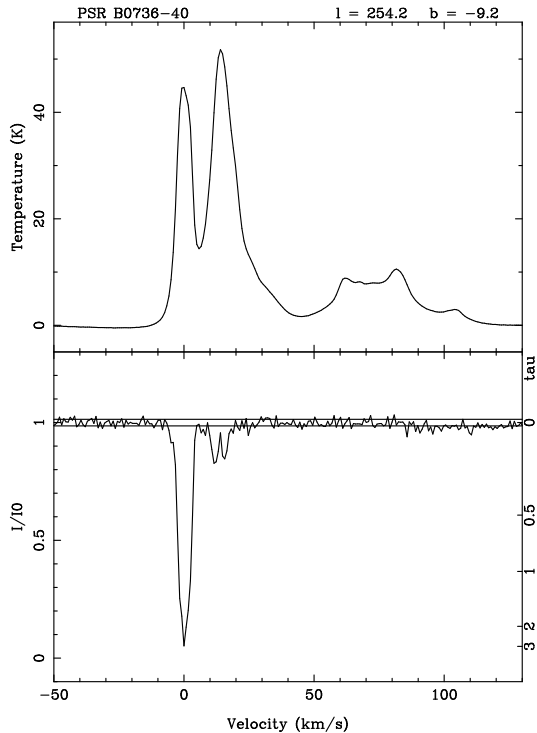}}
  \caption{\HI\ emission (top) and absorption (bottom) spectra 
for PSR B0736--40 from data taken in 2000 August smoothed to 
a velocity resolution of 1.0 \kms. The two horizontal lines on
the lower panel represent the 2-$\sigma$ errors.}
\label{0738_1}
\end{figure}
The data consist of a spectrum for each phase bin for
each of two polarizations for each 10~s of data. These `sub-integrations'
are collected together to form the 45~min integration.
First, the total power (zero lag) from each of the 10~s sub-integrations is
examined and points significantly
higher than the mean are assumed to be affected by
interference and are flagged. Calibration is then performed
on each phase bin and each polarization independently by the total power in
the spectrum.

The pulse profile is then formed
for each polarization by collapsing all time and frequency information.
This yields the location of the `off-pulse' and `on-pulse' bins.
Once these are known, a reference spectrum is
formed by summing all the `off-pulse' bins together over the
entire time span and both polarizations. Bandpass fitting is then
performed on this spectrum (excluding the \HI\ emission channels) with a
polynomial function usually of order 2 or 3.

An on-pulse spectrum for each polarization is then created from 
the phase bin containing the pulse. The two polarizations are kept
separate for the on-pulse measurements: pulsars are highly polarized
and so different power is seen in the two polarization channels.
The absorption spectrum is then created by multiplying this on-pulse
spectrum by its total power and subtracting the reference spectrum
multiplied by its total power. This ensures an off emission baseline
level of 1.0. In practice, however, we found that the baseline was
not perfectly flat after this procedure and so
the resultant absorption spectrum was again bandpass flattened. 
Finally the absorption spectrum is weighted
according to the pulsar power. This allows for observations made
for the different polarization channels and
at different times to be summed together whilst maximising signal
to noise.

The formation of `ghost' absorption is a potential source of error,
especially when observing strong pulsars. This is caused by the
fact that the power from each single pulse is not measured, rather
an average power over the 10-s subintegration is measured. Therefore
if the pulsar is strong (compared to the system temperature) and variable within
the subintegration length, the calibration is not perfect and some
of the emission spectrum leaks into the absorption spectrum \cite{wrb80}.
This can be removed by a least squares fitting process and is discussed
for each pulsar separately in Section 5.

When assessing the differences between spectra from different epochs,
it is vital to take into account the increase in the system temperature
due to the \HI\ contribution. Off-line, the measured rms in the 
spectrum is a result of receiver noise (22 K) and the sky background
(2-15 K). The sky background was computed from the values in
Haslam et al. (1982)\nocite{hssw82} at 408 MHz scaled to 1.4 GHz assuming
a spectral index of --2.6.
On-line, the rms is increased by the \HI\ contribution, which
can have a brightness temperature of up to 120 K. Therefore, the rms 
on the line can be up to 5 times as high as off the line.
The emission spectra for the pulsars have been scaled to fit the
Kerr et al. (1986)\nocite{kbjk86} observations
at the appropriate Galactic coordinates.

\section{Results on individual sources}
\subsection{PSR B0736--40}
\begin{figure}
  \centerline{\psfig{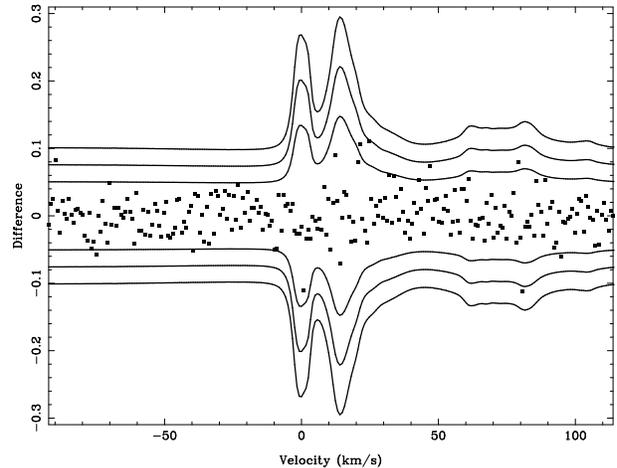}}
  \caption{Difference between the \HI\ absorption spectra of PSR B0736--40
taken in 1998 September and 2000 August at a velocity resolution
of 1.0 \kms. The 1-$\sigma$ level away from the \HI\ signal is 0.025.
Contours shown are 2, 3 and 4-$\sigma$ where the \HI\ temperature in a given
channel has been taken into account.}
\label{0738_2}
\end{figure}
Figure~\ref{0738_1} shows the \HI\ emission and absorption spectra for
PSR B0736--40 from data taken in 2000 August.
The absorption spectrum has been smoothed to yield
an effective velocity resolution of 1.0 \kms\ similar to that
obtained in 1994 \cite{jkww96}.
A comparison between the 1994 and 2000 spectra shows that the
absorption depth was significantly deeper against both emission features
in the 1994 data. In hindsight, we believe that `ghosting' of 
the emission spectrum into the absorption spectrum was taking 
place in 1994 and was not fully corrected for. 
The observation time in 1994 was only 90 mins
compared to 1350 min in 2000. The resultant difference spectrum (not shown)
is dominated by the noise from the earlier epoch and no 
significant changes are seen after the ghosting has been corrected for.

The difference between the spectra taken in 1998 September and 2000 August
is shown in Figure~\ref{0738_2} smoothed to a velocity resolution
of 1.0 \kms.  There are no obvious features in this
difference plot and the distribution of the points are consistent
with Gaussian noise.
The 2-$\sigma$ upper limit is $\sim$0.12 on the line itself
which is somewhat higher than the changes detected in the
pulsars observed by Frail et al. (1994).

\subsection{PSR B1451--68}
\begin{figure}
  \centerline{\psfig{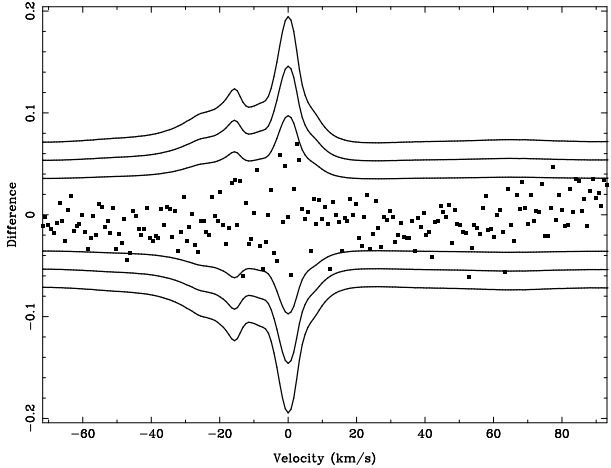}}
  \caption{Difference between the \HI\ absorption spectra in PSR B1451--68
taken in 1998 March and 2000 August at a velocity resolution of 1.0 \kms.
The 1-$\sigma$ level away from the \HI\ signal is 0.018.
Contours shown are 2, 3 and 4-$\sigma$ where the \HI\ temperature in a given
channel has been taken into account.}
\label{1451_1}
\end{figure}

The \HI\ absorption spectrum for this pulsar consists of a single asymmetric
feature peaking at a velocity near 0 \kms\ and 
extending to 6 \kms\ \cite{kjww95}.
The difference between spectra taken in 1998 March and 2000 August
is shown in Figure~\ref{1451_1} after smoothing to a velocity
resolution of 1.0 \kms.
We see no evidence for any significant changes in the absorption spectrum.
Similar results are obtained 
between the 1998 September and 2000 August epochs.
Again, comparison with data taken in 1994 is less interesting as the
on-source time was less than half that obtained for the later data.

\subsection{PSR B1641--45}
\begin{figure}
  \centerline{\psfig{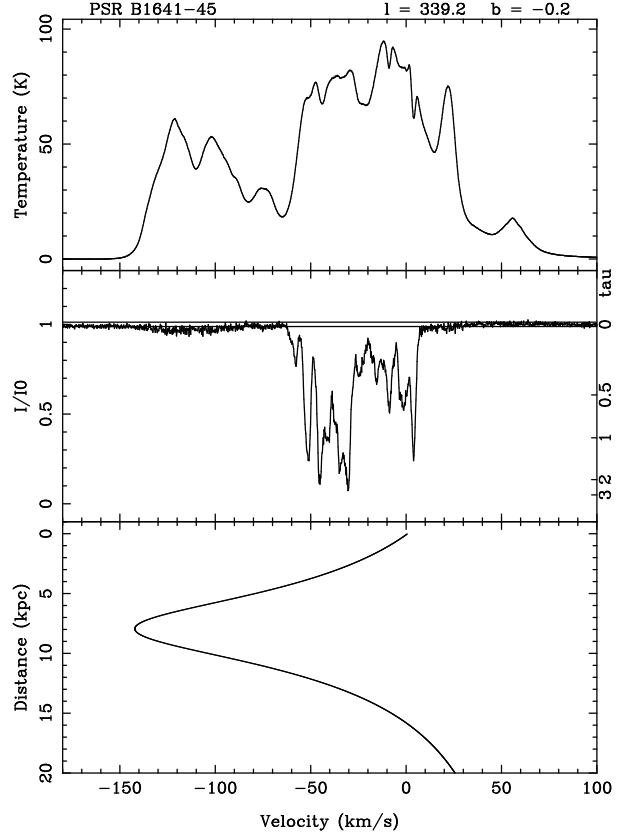}}
  \caption{\HI\ emission (top) and absorption (middle) spectra 
for PSR B1641--45 from data taken in 2000 August at a velocity 
resolution of 0.25 \kms.  The bottom panel shows the rotation curve
appropriate for this Galactic longitude. The shallow dip in 
the absorption spectrum at velocities more negative than --70 \kms\ is 
caused by `ghosting' of the emission spectrum.}
\label{1644_1}
\end{figure}

Figure~\ref{1644_1} shows the \HI\ emission and absorption 
spectra for PSR B1641--45 from data obtained in 2000 Aug.
The absorption spectrum is very similar to that 
obtained at the VLA by Frail et al. (1991)\nocite{fchw91}
although our resolution of 0.25 \kms\ is significantly higher than theirs
(2.5 \kms).
At positive velocities, \HI\ absorption is seen at 4 \kms, presumably
against local gas, but is not seen at 22 \kms against gas located
beyond the solar circle. At negative velocities
the last feature at which absorption is detected is at --58 \kms.
It is apparent from Fig.~\ref{1644_1} that `ghost' absorption is seen
against emission features more negative than --70 \kms\ and also
against the emission at +22 \kms. We removed the effect of this ghosting
by fitting the `ghost' absorption with a fraction of the emission using
a least squares technique. The derived ghosting was at the 5\% level.
\begin{figure}
  \centerline{\psfig{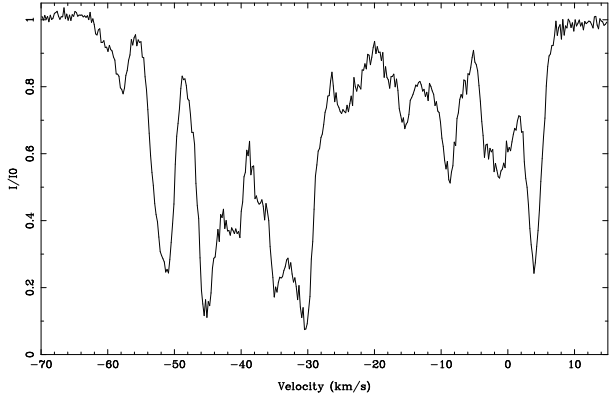}}
  \caption{Detail of the \HI\ absorption spectrum from PSR B1641--45. The
velocity resolution is 0.25 \kms.}
\label{1644_2}
\end{figure}
\begin{figure}
  \centerline{\psfig{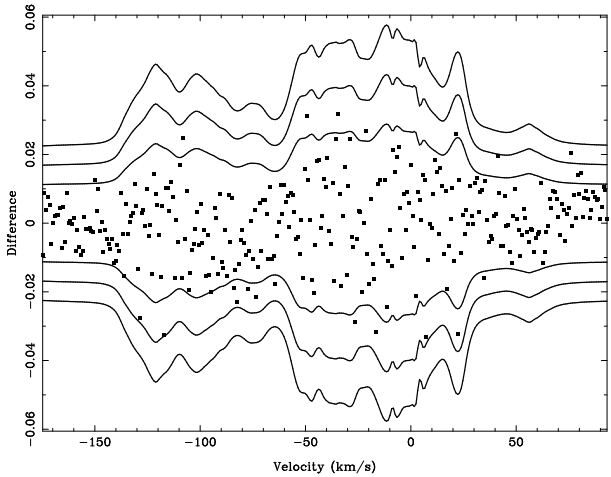}}
  \caption{Difference between the \HI\ absorption spectra of PSR B1641--45
taken in 1998 March and 2000 August at a velocity resolution of 1.0 \kms.
The 1-$\sigma$ level away from the \HI\ signal is 0.0055.
Contours shown are 2, 3 and 4-$\sigma$ where the \HI\ temperature in a given
channel has been taken into account.}
\label{1644_3}
\end{figure}

The \HI\ absorption spectrum is shown in 
more detail in Figure~\ref{1644_2} where
a plethora of narrow lines can be seen which make this pulsar ideal
for a study of opacity variations. At least 12 individual lines are
seen but the full-width half-maxima (FWHM) of the features are not less than 
0.8 \kms\ and are typically 1.2 to 2.0 \kms.
This is similar to results found in other high 
resolution \HI\ studies \cite{cro81} which determined that 
the distribution of the FWHM of absorbing features peaks at 1.8 \kms.

In order to increase the sensitivity to any difference feature
we reduced the velocity resolution of the data from 0.25 to 1.0 \kms\
to better match the observed linewidths.
Figure~\ref{1644_3} shows the difference between the
spectra taken in 1998 March and 2000 August. The significance of the
points has been computed assuming a receiver temperature of 22~K,
a sky temperature of 14~K and the known contribution of the \HI\
in the individual channels combined with the integration time on source.
Contours are shown at the 2, 3 and 4-$\sigma$ levels.
Only one data point has a significance greater than 3-$\sigma$ and
the statistics are consistent with a Gaussian noise distribution.
We therefore conclude that there are no significant changes in 
the \HI\ absorption to any of the 12 individual clouds at 
a 2-$\sigma$ level of 0.022 (on the line). In terms of the difference
spectrum, this sensitivity is better than 5 of the 6 pulsars
in the Frail et al. (1994) survey and roughly on a par with the
sensitivity achieved for PSR B1929+10.

Two other absorption spectra for this pulsar exist at a velocity
resolution of $\sim$2--3~\kms\ taken in 1975 (Figure 8 in
Ables \& Manchester 1976\nocite{am76}) and 1988 (Figure 10 in
Frail et al. 1991\nocite{fchw91}). We have averaged our
data to a similar resolution. Visual inspection of the three spectra
shows that they are very similar in the number and depth of
individual features, with the only possible difference being in
the absorption feature at 4 \kms.

\subsection{PSR B1557--50}
\begin{figure}
  \centerline{\psfig{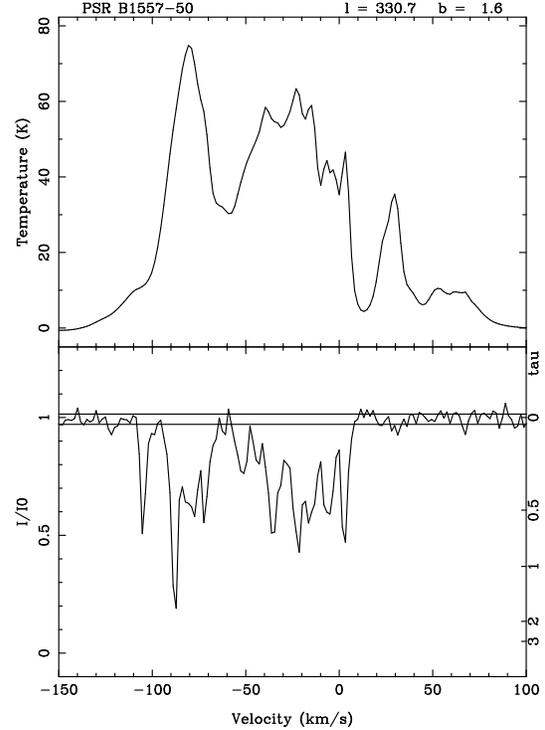}}
  \caption{\HI\ emission and absorption spectra of PSR B1557--50
from 2000 August at a velocity resolution of 2.0 \kms. The two 
horizontal lines on the lower panel represent the 2-$\sigma$ errors.}
\label{1557_2}
\end{figure}
This pulsar is different to the other three under consideration here
in that it is considerably weaker. It was therefore observed with
2.0 \kms\ velocity resolution to match the observations made in 1994.
The resultant absorption spectrum from the 2000 data is shown in
Figure~\ref{1557_2} and can be compared with the 1994 data as shown
in Figure 1 from Johnston et al. (2001).

\begin{figure}
  \centerline{\psfig{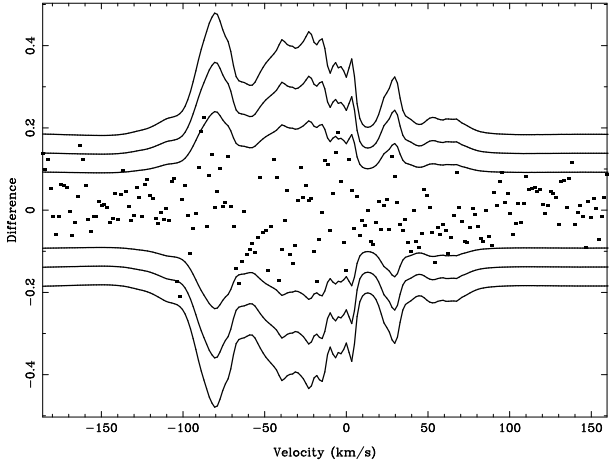}}
  \caption{Difference between the \HI\ absorption spectra of PSR B1557--50
taken in 1994 October and 2000 August at a velocity resolution of 2.0 \kms.
The 1-$\sigma$ level away from the \HI\ signal is 0.05.
Contours are 2, 3 and 4-$\sigma$ where the \HI\ temperature in a given
channel has been taken into account.}
\label{1557_1}
\end{figure}

The difference spectrum between the two data sets are shown is
Figure~\ref{1557_1}.
Very few points lie outside the 2-$\sigma$ contours
as expected from Gaussian noise. However, we note that exactly at
the location of the absorption feature against the cold cloud 
(at --110 \kms) there
are two consecutive outliers. These show that the optical depth in
1994 was similar to that in 2000 August except the line has become
significantly narrower over the 6 year time span.

\begin{figure}
  \centerline{\psfig{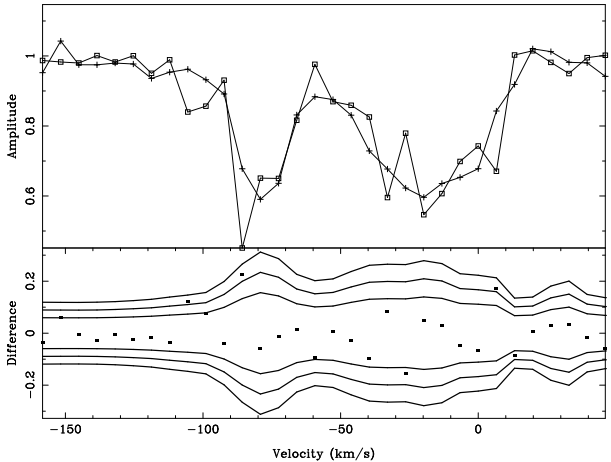}}
  \caption{\HI\ Absorption spectra for PSR B1557--50 from 
1980 (Manchester et al. 1981; plus symbols) and 2000 August (square symbols)
at the same velocity resolution of 7.3 \kms\ (top plot).
The bottom panel shows their difference
with contours at 2, 3 and 4-$\sigma$ significance.}
\label{1557_3}
\end{figure}

\HI\ absorption measurements towards this pulsar were carried 
out in the late 1970s by Ables \& Manchester (1976)\nocite{am76}
and Manchester et al. (1981)\nocite{mwm81}.
Deshpande et al. (1992)\nocite{dmra92} compared these two spectra and 
noted a possible change in the absorption profile near --40 \kms.
The velocity resolution of the earlier observations is a factor of
4 worse than our 1994 and 2000 observations. We therefore box-car 
averaged our data to achieve a similar resolution to the 
Manchester et al. (1981) observations and compare both in
Figure~\ref{1557_3}. To produce the difference
plot we assume the sensitivity was a factor of 3 worse in the
Manchester et al. (1981) observations
(the system temperature is a factor of 2 higher
and the observation time a factor 2.5 less). Differences are clearly
apparent. The deep absorption feature at --110 \kms\ is not
present in the earlier observations and there are additional differences
in both the features at --80 and --40 \kms.

\section{Discussion}
No significant differences were found in the \HI\
absorption spectra of PSRs B0736--40, B1451--68 or B1641--45 over
a 2.5 yr timescale. At full velocity resolution, 
our difference spectra have rms levels which are
not as good as those achieved by Frail et al. (1994) and
examination of their spectra show that the differences they detected
were below our sensitivity limit.
However, smoothing the data to a resolution equivalent to 
that of the measured line widths ($\sim$1.0 \kms)
brings us within a factor of 2 in sensitivity for PSRs B0736--40 and B1451--68.
For PSR B1641--45 our sensitivity level is lower than any
of the differences claimed by Frail et al. (1994) and our non-detection
is therefore a significant result.
The upper limit we can put on \HI\ column density variations to each
of the 12 individual absorbing features along the line of
sight to this pulsar, assuming a spin temperature of 50~K,
is $10^{19}$~cm$^{-2}$.
This is significantly less than the changes seen in VLBI
\HI\ and molecular line observations and comparable to the smallest
changes seen in pulsar absorption spectra.

We also raise some issues concerning the interpretation of the \HI\ spectra
in Frail et al. (1994). First, the difference spectra (see Fig.~3 
in their paper) are shown as absolute differences rather than
centered at zero; this has an effect on the noise statistics.
Secondly, it is clear that
because the temperature increases on the \HI\ line so does the noise
and although they state in the text that this is taken into
account the lack of significance contours on the figure make it
difficult to interpret. For PSR B0540+23, for example, the 
noise must be at least
a factor of 3 higher on the line (emission temperature of 120 K) compared
to off the line. The observed difference spectra are indeed significantly 
noisier under the line but this is interpreted as 
real differences in the discussion. For
PSR B1929+10, variations against the baseline seem just 
as large as against the line but this point is not discussed.
Finally, it appears as if the pulsars with the best sensitivity in turn
show the smallest difference variations.
We note in mitigation, however, that their technique of 
using the equivalent width of the lines is more robust than the 
simple difference figures.

As discussed in Johnston et al. (2001)\nocite{jkww01},
the \HI\ absorption feature at --110 \kms\ in PSR B1557--50
is rather unusual. It has an apparent \HI\ emission temperature of only 20~K,
and one does not expect to see significant optical depth 
against such features \cite{fw90}.
However, its line width is not particularly narrow; the
FWHM of the line is $\sim$3~\kms,
indicative of a much warmer object if interpreted as thermal broadening.
It therefore seems likely that the cloud has a 
spin temperature in the range 50--100~K but that beam dilution causes
a lower temperature to be observed.
At this spin temperature and an equivalent width of $\sim$3~\kms\
the column density of neutral gas is $\sim$4$\times$10$^{20}$~cm$^{-2}$.
We can make some estimate of the size of the absorbing feature
because the line-of-sight to the pulsar did not go through the cloud
in 1980, did so in 1994 and appears to be weakening again in 2000.
Therefore the pulsar takes roughly 12 years to cross the cloud.
The absorption feature is located at 6.4 kpc and the pulsar located beyond this,
although the dispersion measure derived distance would suggest that
the cloud-pulsar distance is likely small. We therefore ignore geometrical
effects. Assuming a velocity of 400 \kms\ for the pulsar (likely correct
within a factor of 2) the pulsar then moves 1000 AU in that time.
Assuming the cloud is spherical, this yields a density of
$2.6\times$10$^{4}$~cm$^{-3}$. This is similar to the values obtained
from other observations as outlined in Section 1.
Deshpande (2000)\nocite{desh00}, however, has shown that these results
can instead be interpreted as much smaller density variations on
larger scales when {\it all\/} contributing scales are taken into
account, rather than simply the transverse scale.
In his picture, it is not surprising that the pulsar
with the longest time interval shows the largest variations in optical
depth - indeed one expects that as the time interval increases, larger and
larger deviations in optical depth should eventually be detected.
The non-detection of optical depth variations at the 
$<0.2$ level in the three 
pulsars with a 2.5 yr time interval are also consistent with his
derived structure function.

\section{Conclusions}
Although overwhelming evidence exists for small-scale
fluctuations in the interstellar medium,
observations of three southern pulsars over a 2.5 yr time span have failed
to reveal any significant differences in their \HI\ absorption spectra.
The overall sensitivity of our high resolution observations is
somewhat less than that of 
Frail et al. (1994), but we have correctly handled the variation
in noise level across the \HI\ emission. After smoothing the data
to a velocity resolution of 1.0 \kms, the upper limit of 
$10^{19}$~cm$^{-2}$ in the change in column density of the cold, neutral
gas towards PSR B1641--45 is significantly less than changes seen
in the direction of other pulsars. \HI\ absorption spectra of
PSR B1557--50 taken in 1994 and 2000 clearly show differences
and these are even more pronounced when comparison with spectra taken in
the late 1970s is made. The differences are most evident in a high optical
depth feature against an emission feature with an
apparent temperature of only 20~K.

\section*{Acknowledgments}
We thank J. Weisberg for encouraging discussions and D. Mitra
for helpful comments on the text.
The Australia Telescope is funded by the Commonwealth of Australia
for operation as a National Facility managed by the CSIRO.

\bibliographystyle{mn}
\bibliography{modrefs,psrrefs,crossrefs}

\begin{thebibliography}{{Manchester, Wellington \& McCulloch }{1981}}

\bibitem[\protect\citename{Ables \& Manchester }{1976}]{am76}
Ables~J.~G., Manchester~R.~N., 1976, A\&A, 50, 177

\bibitem[\protect\citename{Bailes {\rm et~al. }}{1990}]{bmk+90a}
Bailes~M., Manchester~R.~N., Kesteven~M.~J., Norris~R.~P., Reynolds~J.~E.,
  1990, Nat, 343, 240

\bibitem[\protect\citename{Crovisier }{1981}]{cro81}
Crovisier~J., 1981, A\&A, 94, 162

\bibitem[\protect\citename{Davis, Diamond \& Goss }{1996}]{ddg96}
Davis~R.~J., Diamond~P.~J., Goss~W.~M., 1996, MNRAS, 283, 1105

\bibitem[\protect\citename{Deshpande }{2000}]{desh00}
Deshpande~A.~A., 2000, MNRAS, 317, 199

\bibitem[\protect\citename{Deshpande, Dwarakanath \& Goss }{2000}]{ddg00}
Deshpande~A.~A., Dwarakanath~K.~S., Goss~W.~M., 2000, ApJ, 543, 227

\bibitem[\protect\citename{Deshpande {\rm et~al. }}{1992}]{dmra92}
Deshpande~A.~A., McCulloch~P.~M., Radhakrishnan~V., Anantharamaiah~K.~R., 1992,
  MNRAS, 258, 19P

\bibitem[\protect\citename{Diamond {\rm et~al. }}{1989}]{dgr+89}
Diamond~P.~J., Goss~W.~M., Romney~J.~D., Booth~R.~S., Kalberla~P.~M.~W.,
  Mebold~U., 1989, ApJ, 347, 302

\bibitem[\protect\citename{Dickey \& Lockman }{1990}]{dl90}
Dickey~J.~M., Lockman~F.~J., 1990, ARA\&A, 28, 215

\bibitem[\protect\citename{Downs \& Reichley }{1983}]{dr83}
Downs~G.~S., Reichley~P.~E., 1983, ApJS, 53, 169

\bibitem[\protect\citename{Faison \& Goss }{2001}]{fg01}
Faison~M.~D., Goss~W.~M., 2001, ApJ, 121, 2706

\bibitem[\protect\citename{Faison {\rm et~al. }}{1998}]{fgdt98}
Faison~M.~D., Goss~W.~M., Diamond~P.~J., Taylor~G.~B., 1998, AJ, 116, 2916

\bibitem[\protect\citename{Fomalont {\rm et~al. }}{1992}]{fgl+92}
Fomalont~E.~B., Goss~W.~M., Lyne~A.~G., Manchester~R.~N., Justtanont~K., 1992,
  MNRAS, 258, 497

\bibitem[\protect\citename{Frail \& Weisberg }{1990}]{fw90}
Frail~D.~A., Weisberg~J.~M., 1990, AJ, 100, 743

\bibitem[\protect\citename{Frail {\rm et~al. }}{1991}]{fchw91}
Frail~D.~A., Cordes~J.~M., Hankins~T.~H., Weisberg~J.~M., 1991, ApJ, 382, 168

\bibitem[\protect\citename{Frail {\rm et~al. }}{1994}]{fwcm94}
Frail~D.~A., Weisberg~J.~M., Cordes~J.~M., Mathers~C., 1994, ApJ, 436, 144

\bibitem[\protect\citename{Gwinn }{2001}]{gwi01}
Gwinn~C.~R., 2001, ApJ, 561, 815

\bibitem[\protect\citename{Haslam {\rm et~al. }}{1982}]{hssw82}
Haslam~C. G.~T., Salter~C.~J., Stoffel~H., Wilson~W.~E., 1982, A\&AS, 47, 1

\bibitem[\protect\citename{Heiles }{1997}]{hei97}
Heiles~C., 1997, ApJ, 481, 193

\bibitem[\protect\citename{Johnston {\rm et~al. }}{1996}]{jkww96}
Johnston~S., Koribalski~B.~S., Weisberg~J.~M., Wilson~W., 1996, MNRAS, 279, 661

\bibitem[\protect\citename{Johnston {\rm et~al. }}{2001}]{jkww01}
Johnston~S., Koribalski~B.~S., Weisberg~J.~M., Wilson~W., 2001, MNRAS, 322, 715

\bibitem[\protect\citename{Kerr {\rm et~al. }}{1986}]{kbjk86}
Kerr~F.~J., Bowers~P.~F., Jackson~P.~D., Kerr~M., 1986, A\&AS, 66, 373

\bibitem[\protect\citename{Koribalski {\rm et~al. }}{1995}]{kjww95}
Koribalski~B.~S., Johnston~S., Weisberg~J.~M., Wilson~W., 1995, ApJ, 441, 756

\bibitem[\protect\citename{Lauroesch, Meyer \& Blades }{2000}]{lmb00}
Lauroesch~J.~T., Meyer~D.~M., Blades~J.~C., 2000, ApJ, 543, L43

\bibitem[\protect\citename{Manchester, Wellington \& McCulloch }{1981}]{mwm81}
Manchester~R.~N., Wellington~K.~J., McCulloch~P.~M., 1981, in Sieber~W.,
  Wielebinski~R., eds, Pulsars, {IAU} {S}ymposium 95.
\newblock Reidel, Dordrecht, p.~445

\bibitem[\protect\citename{Meyer \& Lauroesch }{1999}]{ml99}
Meyer~D.~M., Lauroesch~J.~T., 1999, ApJ, 520, L103

\bibitem[\protect\citename{Mitra \& Ramachandran }{2001}]{mr01}
Mitra~D., Ramachandran~R., 2001, A\&A, 370, 586

\bibitem[\protect\citename{Moore \& Marscher }{1995}]{mm95}
Moore~E.~M., Marscher~A.~P., 1995, ApJ, 452, 671

\bibitem[\protect\citename{Staveley-Smith {\rm et~al. }}{1996}]{swb+96}
Staveley-Smith~L. {\rm et~al.}, 1996, Proc.\,Astr.\,Soc.\,Aust., 13, 243

\bibitem[\protect\citename{Taylor, Manchester \& Lyne }{1993}]{tml93}
Taylor~J.~H., Manchester~R.~N., Lyne~A.~G., 1993, ApJS, 88, 529

\bibitem[\protect\citename{Watson \& Meyer }{1996}]{wm96}
Watson~J.~K., Meyer~D.~M., 1996, ApJ, 473, L127

\bibitem[\protect\citename{Weisberg, Rankin \& Boriakoff }{1980}]{wrb80}
Weisberg~J.~M., Rankin~J.~M., Boriakoff~V., 1980, A\&A, 88, 84

\end{thebibliography}
\end{document}